\documentclass[aps,reprint,superscriptaddress,showkeys,onecolumn,11pt]{revtex4-2}

\usepackage{color}
\usepackage[version=3]{mhchem} 
\usepackage{gensymb}
\usepackage{amssymb}
\usepackage{upgreek}
\usepackage{mathtools}
\usepackage{comment}
\usepackage[dvipsnames]{xcolor}
\usepackage[T1]{fontenc}

\begin{document}

\title {Quantum Emitters in Rhombohedral Boron Nitride}

\author{Angus Gale}
\affiliation{School of Mathematical and Physical Sciences, Faculty of Science, University of Technology Sydney, Ultimo, New South Wales 2007, Australia}

\author{Mehran Kianinia}
\affiliation{School of Mathematical and Physical Sciences, Faculty of Science, University of Technology Sydney, Ultimo, New South Wales 2007, Australia}
\affiliation{ARC Centre of Excellence for Transformative Meta-Optical Systems (TMOS), University of Technology Sydney, Ultimo, New South Wales 2007, Australia}

\author{Jake Horder}
\affiliation{School of Mathematical and Physical Sciences, Faculty of Science, University of Technology Sydney, Ultimo, New South Wales 2007, Australia}

\author{Connor Tweedie}
\affiliation{School of Mathematics and Physics, The University of Queensland, Brisbane, QLD 4072, Australia}

\author{Mridul Singhal}
\affiliation{School of Mathematics and Physics, The University of Queensland, Brisbane, QLD 4072, Australia}

\author{Dominic Scognamiglio}
\affiliation{School of Mathematical and Physical Sciences, Faculty of Science, University of Technology Sydney, Ultimo, New South Wales 2007, Australia}

\author{Jiajie Qi}
\affiliation{State Key Laboratory for Mesoscopic Physics, School of Physics, Peking University, Beijing 100871, China}

\author{Kaihui Liu}
\affiliation{State Key Laboratory for Mesoscopic Physics, School of Physics, Peking University, Beijing 100871, China}
\affiliation{Songshan Lake Materials Laboratory, Dongguan 523808, China}

\author{Carla Verdi}
\affiliation{School of Mathematics and Physics, The University of Queensland, Brisbane, QLD 4072, Australia}

\author{Igor Aharonovich}
\email[]{igor.aharonovich@uts.edu.au}
\affiliation{School of Mathematical and Physical Sciences, Faculty of Science, University of Technology Sydney, Ultimo, New South Wales 2007, Australia}
\affiliation{ARC Centre of Excellence for Transformative Meta-Optical Systems (TMOS), University of Technology Sydney, Ultimo, New South Wales 2007, Australia}

\author{Milos Toth}
\email[]{milos.toth@uts.edu.au}
\affiliation{School of Mathematical and Physical Sciences, Faculty of Science, University of Technology Sydney, Ultimo, New South Wales 2007, Australia}
\affiliation{ARC Centre of Excellence for Transformative Meta-Optical Systems (TMOS), University of Technology Sydney, Ultimo, New South Wales 2007, Australia}

\date{\today}

\begin{abstract}
\vspace{1cm}
Rhombohedral boron nitride (rBN) is an emerging wide-bandgap van der Waals (vdW) material that combines strong second-order nonlinear optical properties with the structural flexibility of layered 2D systems. Here we show that rBN hosts optically-addressable spin defects and single-photon emitters (SPEs). Both are fabricated deterministically, using site-specific techniques, and are compared to their analogues in hexagonal boron nitride (hBN). Emission spectra in hBN and rBN are compared, and computational models of defects in hBN and rBN are used to elucidate the debated atomic structure of the B-center SPE in BN. Our results establish rBN as a monolithic vdW platform that uniquely combines second-order nonlinear optical properties, optically addressable spin defects, and high-quality SPEs, opening new possibilities for integrated quantum and nonlinear photonics.
\vspace{1cm}
\end{abstract}


\maketitle 

\clearpage

\section{Introduction}

Solid-state defects in Van der Waals (vdW) materials have great potential for use in next-generation quantum technologies. The defects can be employed as quantum light sources, which form the basis of secure communication and photonic quantum computation schemes~\cite{Turunen:2022, Liu:2019}. Hexagonal boron nitride (hBN) is a compelling wide bandgap vdW material, unique in that it hosts high quality single photon emitters (SPEs) and optically-addressable spin defects~\cite{Vaidya:2023, Aharonovich:2022}. They have been incorporated in devices such as waveguides and cavities that underpin monolithic integrated quantum photonics~\cite{Gerard:2023, Nonahal:2023, Zotev:2023, Kuhner:2023}. 

Rhombohedral boron nitride (rBN) is analogous to hBN in that it consists of atomic sheets of B and N atoms arranged in a 2D hexagonal lattice. The two polytypes differ in that the atomic sheets in hBN are ordered in an AA' stacking sequence, whilst rBN has the ABC stacking~\cite{Moret:2021} shown in Figure \ref{Fig_Overview}a. A consequence of AA' stacking is that, whilst monolayer BN exhibits second-order optical nonlinearities, the centrosymmetric AA' structure of hBN inhibits them in bulk hBN~\cite{Qi:2024}. Conversely, the ABC stacking of rBN breaks inversion symmetry, giving rise to highly-efficient nonlinear optical processes that enable second-harmonic generation (SHG) and parametric down-conversion~\cite{Liang:2025}. Indeed, an SHG efficiency of 1\% has recently been reported in bulk rBN, along with methods for large-scale growth of rBN~\cite{Qi:2024, Wang:2024}. 

Here, we introduce optically-addressable spin defects and SPEs in rBN. We show that rBN can host negatively charged boron vacancy (\ce{V_B-}) ensembles and B-center SPEs, and compare them to their counterparts in hBN~\cite{Clua-Provost:2023, Fournier:2021}. We focus on these two defects because \ce{V_B-} is the most-studied spin-defect in hBN, and the B-center is presently considered the most promising hBN SPE for on-chip quantum photonics~\cite{Aharonovich:2022}. We confirm the \ce{V_B-} spin defect emission unambiguously by optically detected magnetic resonance (ODMR) spectroscopy, and perform cryogenic photoluminescence (PL) spectroscopy of B-centers in rBN and hBN. The spectra are simulated using computational models of defects in hBN and rBN, and used to elucidate the debated~\cite{Zhigulin2023, Ganyecz:2024, Maciaszek:2024} atomic structure of the B-center defect, concluding that an in-plane carbon chain tetramer (C4)~\cite{Maciaszek:2024} is the most likely origin of the B-center emission in hBN and rBN.

Our work establishes rBN as a vdW platform for monolithic integrated photonics. It combines strong second-order nonlinear optical properties, optically addressable spin defects, and high-quality SPEs in a bulk vdW material. Other vdW systems exhibit strong second order nonlinearities only at monolayer or few-layer thicknesses (e.g., hBN, \ce{MoS2}, \ce{WS2}, \ce{MoSe2}, \ce{WSe2}) which are incompatible with monolithic devices such as waveguides and photonic crystal cavities, and/or do not host spin defects and high quality SPEs (e.g., GaSe, SnS, SnSe, 3R-\ce{MoS2})~\cite{Autere:2018, Jie:2015, Moqbel:2024, Mao:2023, Xu:2022}. rBN not only satisfies these criteria simultaneously, it is also optically-transparent across a broad spectral range due to its wide bandgap of $\rm\sim$5.8~eV~\cite{Qi:2024}. It is therefore presently unique as a vdW material and matched only by SiC~\cite{Lukin:2020} as a platform for monolithic integrated quantum and nonlinear photonics, offering both intrinsic second-order optical nonlinearity, spin defects and high-quality SPEs within a single material system.

\section{Results and Discussion}

\begin{figure}[!h]
\includegraphics[width=0.95\linewidth]{"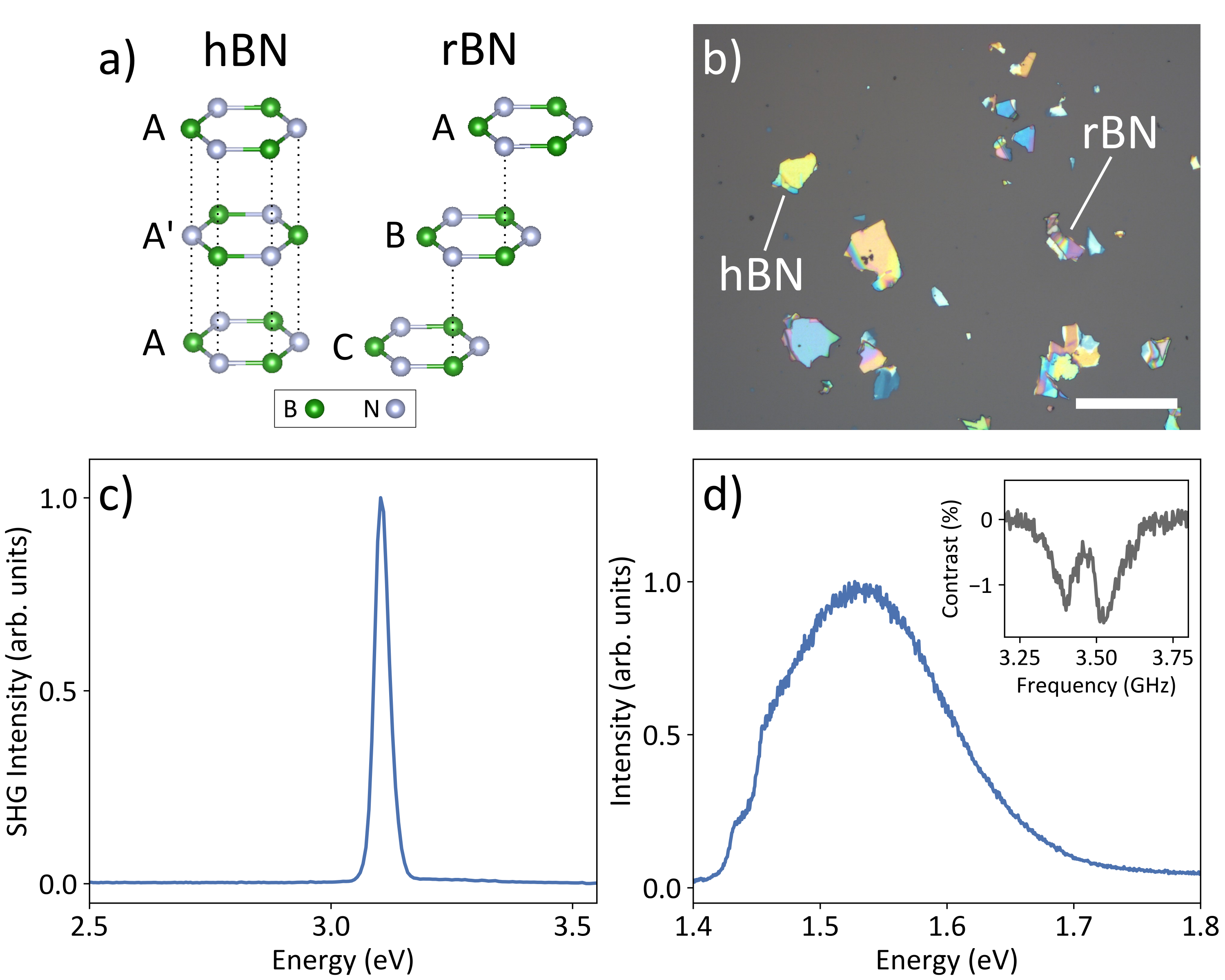"}
\caption{Overview of hBN and rBN crystals, second harmonic generation (SHG) and \ce{V_B-} spin defect in rBN. a) Schematic of hBN and rBN crystal structure and stacking order. b) Optical image of BN flakes exfoliated onto a \ce{Si/SiO2} substrate. The scale bar corresponds to 50~$\rm\mu$m. c) Reflected second harmonic generation spectrum from an exfoliated rBN flake obtained using 1.55~eV (800~nm) excitation. d) Room temperature PL spectrum of an ensemble of \ce{V_B-} defects in an rBN flake. (Inset) ODMR spectrum from the ensemble.}
\label{Fig_Overview}
\end{figure}

The hBN and rBN polytypes differ by the monolayer stacking order, as shown in Figure \ref{Fig_Overview}a. Both polytypes can be easily exfoliated, and the optical image in Figure \ref{Fig_Overview}b shows a number of exfoliated BN flakes transferred to an \ce{Si/SiO2} substrate. When multiple BN polytypes are present, as in this instance, optical microscopy is not sufficient to identify the stacking order. Typically, flake color is the only visible difference optically, which is related to both the substrate and flake thickness. Due to the strong optical non-linear response of rBN, SHG measurements can be used to differentiate the stacking order of BN \cite{Qi:2024}. An example of the SHG response from an exfoliated rBN flake is shown in Figure \ref{Fig_Overview}c using a 1.55~eV (800~nm) excitation laser.

We start by demonstrating that rBN can host \ce{V_B-} spin defects. At present, \ce{V_B-} is the only defect in hBN for which the atomic structure has been unambiguously identified. The defect is ODMR-active, making it an appealing platform for quantum sensing applications~\cite{Healey:2023}. Although single \ce{V_B-} defects have not yet been isolated optically, \ce{V_B-} ensembles can be readily engineered in hBN using site-specific ion beam irradiation~\cite{Sarkar:2024}. Here, an exfoliated rBN flake was irradiated with a 30~keV nitrogen ion beam to generate defects. Figure \ref{Fig_Overview}d shows a room temperature (RT) PL spectrum of the irradiated region, revealing a broad emission centered on $\sim$1.5~eV, typical of the \ce{V_B-} PL spectrum in hBN. A corresponding ODMR spectrum is shown in the inset, confirming unambiguously the presence of the \ce{V_B-} spin sub-levels. As with hBN, \ce{V_B-} in rBN has two distinct dips in the ODMR spectrum centered on $\sim$3.45~GHz, corresponding to the zero-field splitting of the $\rm+1$ and $\rm-1$ spin sublevels. The results in Figure \ref{Fig_Overview} show that rBN can host ODMR-active \ce{V_B-} spin defects.

We note that the PL spectrum of \ce{V_B-} ensembles is dominated by a broad phonon sideband (PSB), with a shape that is inconsistent, likely due background PL. This makes it unsuitable for nuanced studies of changes in spectral shape caused by BN stacking order. However, the spectra of other defects can vary consistently between hBN and rBN.

\begin{figure}[h!]
\includegraphics[width=0.95\linewidth]{"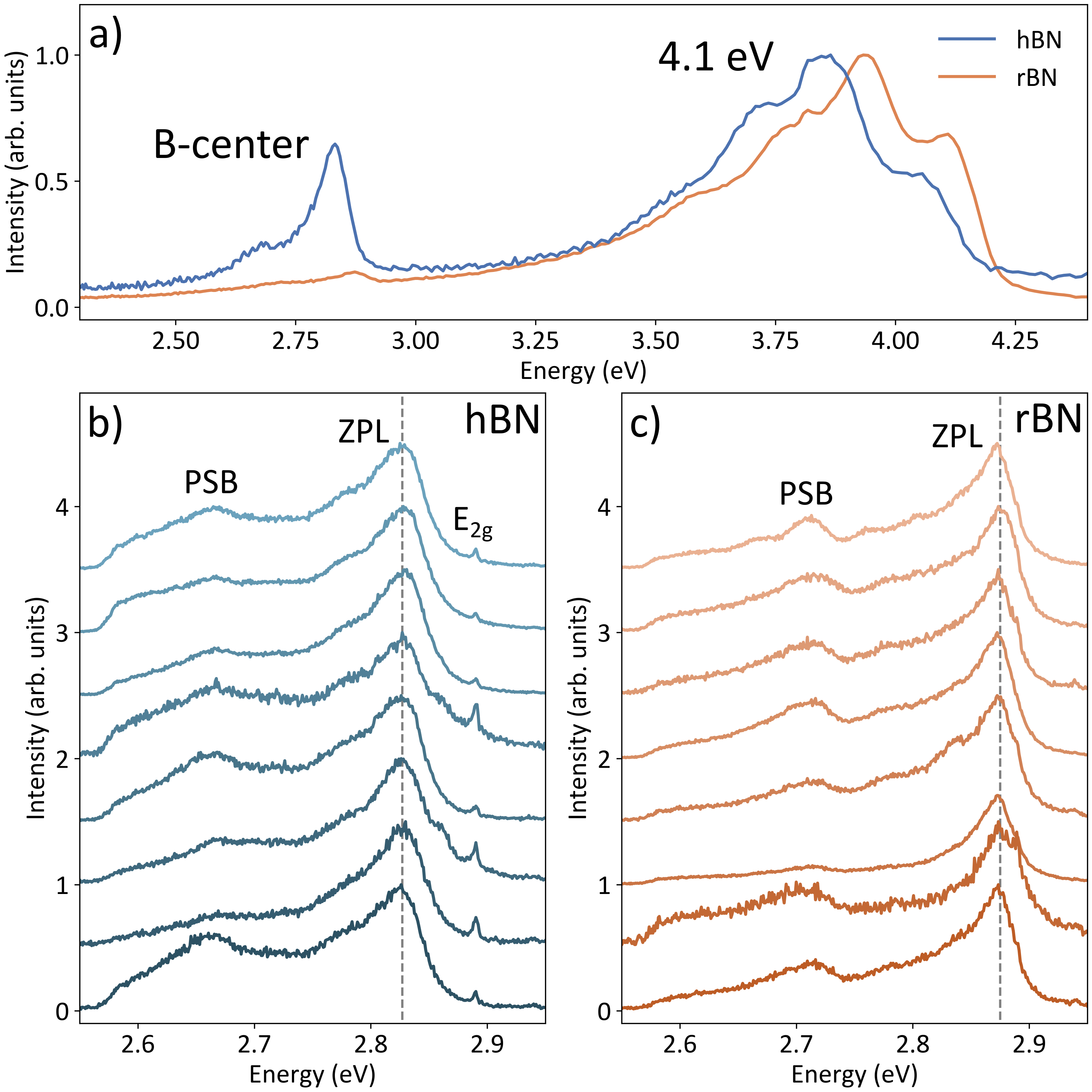"}
\caption{Room temperature emission spectra of hBN and rBN. a) CL spectra from hBN and rBN. Ensemble emissions corresponding to the B-center and 4.1~eV defects are labeled on the plot. b) PL spectra of 8 localized sites with B-center emissions in hBN. The peak at $\rm\sim$2.9~eV corresponds to the $\rm E_{2g}$ BN Raman line. The dip at $\rm\sim$2.55~eV is caused by a band-pass filter in the optical setup. c) PL spectra of 8 localized sites with B-center emissions in rBN. The vertical dashed line corresponds to the ZPL energy. The vertical dashed lines in b and c show the ZPL energy in hBN and rBN, respectively.}
\label{Fig_CLPL}
\end{figure}

Figure \ref{Fig_CLPL}a shows cathodoluminescence (CL) spectra acquired from flakes of hBN and rBN. Each spectrum contains two emissions, labeled `B-center' and `4.1~eV'. The latter corresponds to an ultraviolet (UV) defect ensemble in BN. It has been studied previously in different BN polytypes, revealing shifts in the zero phonon line (ZPL) energy that depend on the BN stacking order~\cite{Iwanski:2024, Rousseau:2022}. Specifically, a 47~meV blue-shift of the emission has been reported in rBN relative to hBN ~\cite{Iwanski:2024}. This shift, observed also in Figure \ref{Fig_CLPL}a, confirms the rhombohedral and hexagonal nature of our rBN and hBN flakes.

The UV defect has been studied extensively in the hBN polytype, where single defects have been isolated and characterized by CL spectroscopy~\cite{Bourrellier:2016}. The UV emitters are, however, not favorable for on-chip quantum applications, and we therefore focus here on the blue B-center emission seen in Figure \ref{Fig_CLPL}a. 

In our experiments, CL was excited using a 3~keV, 26~nA electron beam and, in hBN, it is the electron irradiation itself that generates the emission at $\rm\sim$2.8~eV~\cite{Fournier:2021}. Here, we observed that the beam also activates B-centers in rBN. We note that the relative intensity of the B-center emission seen in Figure \ref{Fig_CLPL}a is greater in hBN than in rBN. This, however, cannot be ascribed to the rBN polytype because the B-center activation rate and emission intensity vary from flake to flake in both polytypes, likely due to variations in local defect structures and densities.

Figure~\ref{Fig_CLPL}b and c are PL spectra showing the B-center emission, obtained from 8 different sites on hBN and rBN flakes, respectively (the flakes were pre-irradiated by the electron beam to activate the emitters). The B-center ZPL in hBN is located consistently at $\rm\sim$2.83~eV, whilst that in rBN is blue-shifted to $\rm\sim$2.87~eV. The sharp peak seen in Figure~\ref{Fig_CLPL}b at $\rm\sim$2.89~eV is the $\rm E_{2g}$ Raman line generated by the 3.06~eV excitation laser (it is not apparent in the rBN spectra due to the blue-shifted B-center emission). 

\begin{figure}[h!]
\includegraphics[width=0.95\linewidth]{"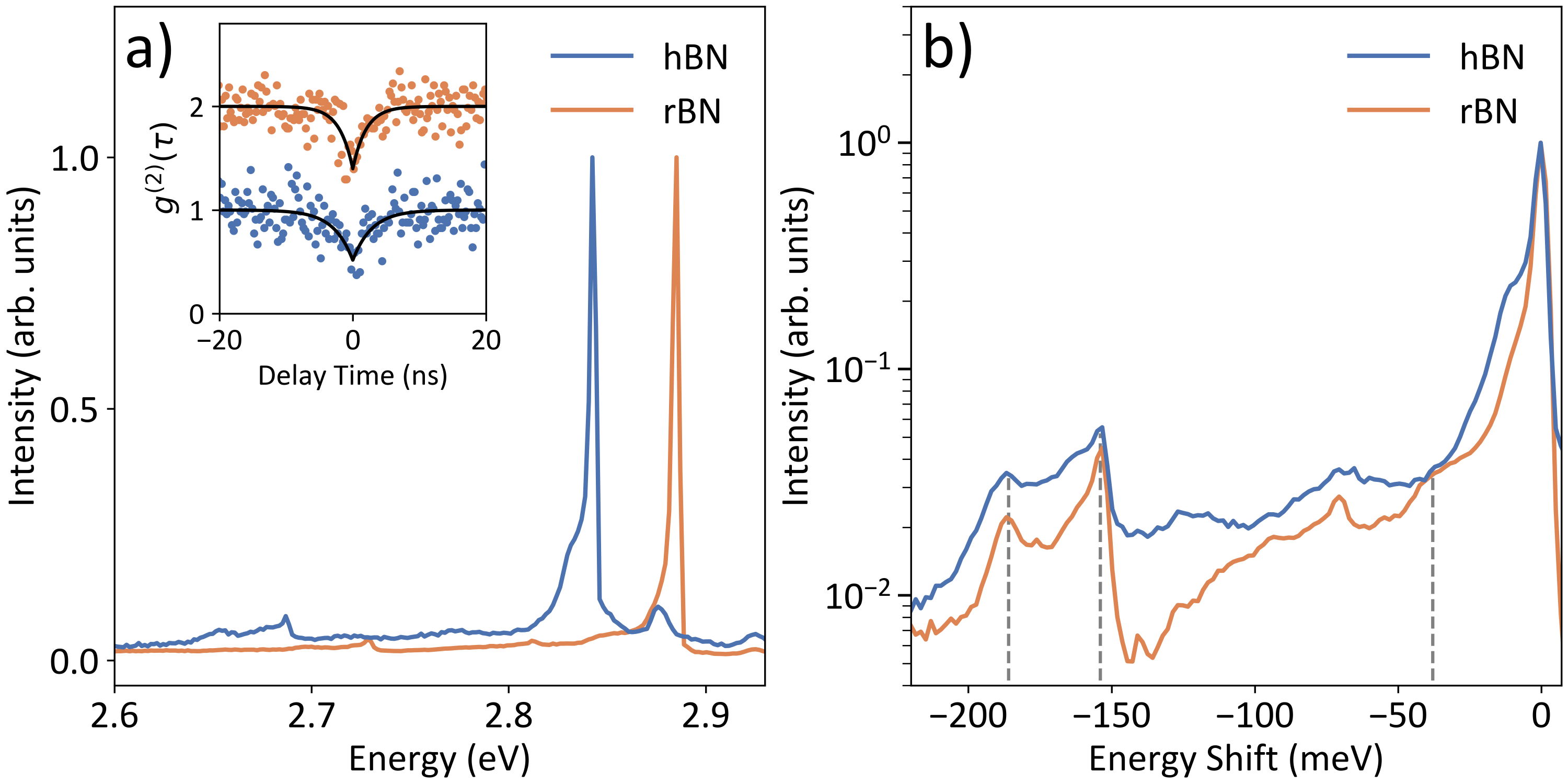"}
\caption{Cryogenic PL spectra of B-centers in hBN and rBN. a) PL spectra at 5~K from single B-centers in hBN and rBN respectively. (Inset) Corresponding autocorrelation measurements from the B-centers in a. The $\rm g^2(0)$ values are 0.4 for rBN and 0.5 for hBN. b) Summed PL spectra at 5~K from B-centers in hBN and rBN. A logarithmic scale is used on the y-axis for clarity in distinguishing the phonon replicas. The vertical dashed lines correspond to LO (-186~meV), TO (-154~meV) and ZA/ZO ($\sim$40~meV) phonons.}
\label{Fig_Cryo}
\end{figure}

Next, we performed cryogenic PL measurements of B-centers to enable a detailed comparison of the emissions in hBN and rBN. Figure~\ref{Fig_Cryo}a shows PL spectra from isolated B-centers in hBN and rBN measured at 5~K. Autocorrelation measurements from each site demonstrate that the the B-center in rBN is a single defect with $\rm g^2(0)=0.4$ (Fig. \ref{Fig_Cryo}a (inset)). The B-center in hBN has similar $\rm g^2(0)$ of 0.5. The narrow ZPLs at 5~K reveal a clear blue shift of 42.5~meV in rBN relative to hBN.

To further compare B-centers in the two BN polytypes, PL spectra were taken from a number of B-center sites in both hBN and rBN. Spectra from each polytype were summed and ZPLs aligned to present the spectra as a function of relative energy shift from the ZPL in Figure~\ref{Fig_Cryo}b (see Figure~S1 for the individual spectra). There are shared spectral features in each polytype including the most prominent peaks at -154~meV and -186~meV, which may be attributed to electron-phonon coupling with the high-energy transverse optical (TO) and longitudinal optical (LO) phonon branches, respectively~\cite{Khatri:2019, Vuong:2016}. The phonon replicas at approximately -70~meV differ slightly in width but are also present in each polytype. However, the replica at $\sim$-40~meV is only present in the rBN spectrum as a broad shoulder with the ZPL. This likely originates from coupling to out-of-plane vibrational modes~\cite{Khatri:2019}. A maximum is present at $\sim$35~meV in the calculated phonon density of states (DOS) for hBN which aligns with the out-of-plane ZA and ZO branches in each polytype~\cite{Zanfrognini:2023, Serrano:2007}, but the broad experimental peak makes it difficult to accurately identify the exact spectral position. The different phonon contributions in the spectra of each polytype are not unexpected, given the effect of stacking order on possible coupling strengths to out-of-plane modes, unlike the higher energy in-plane LO and TO modes. The shoulder at low energies ($\sim$-20~meV) is present only on the spectrum from the hBN flake, but some broadening is also apparent in the rBN spectrum, likely due to a convolution of the $\rm E_{2g}$ Raman line with the ZPL.

In hBN, the B-center quantum emitter has received significant attention, largely because it can be engineered deterministically by electron beam irradiation~\cite{Fournier:2021}, has a highly-consistent ZPL energy~\cite{Fournier:2023}, and it can therefore be coupled to cavities and on-chip optical circuits using methods that have high potential for scalability~\cite{Nonahal:2023, Gerard:2023}. However, despite much research, the atomic structure of the B-center is unknown, and a matter of debate in the literature. 

Our PL data from the hBN and rBN polytypes provides an opportunity to elucidate the atomic structure of the B-center defect. We therefore performed density functional theory (DFT) calculations (see Methods in the Supporting Information) for the most compelling structures proposed to date, namely the out-of-plane carbon split interstitial dimer \ce{C_NC_i}~\cite{Zhigulin2023}, the nitrogen split interstitial \ce{N_i}~\cite{Ganyecz:2024}, and the in-plane carbon chain tetramer C4~\cite{Maciaszek:2024}.

B-centers are expected to possess $D_{3h}$ symmetry and a singlet ground state~\cite{Zhigulin2023}. This rules out the carbon dimer defect \ce{C_NC_i}, as shown by our calculations. In the negative charge state, \ce{C_NC_i^-}, we find that the symmetric $D_{3h}$ dumbbell structure in hBN is metastable and has a spin triplet configuration, while the ground state exhibits lower symmetry in both the spin-triplet and spin-singlet state (see Figure S2 of the Supporting Information). Similarly, the $D_{3h}$ structure is unfavorable in the neutral charge state, \ce{C_NC_i^0}. The rBN polytype yields comparable ground-state geometries.

\begin{figure}[h!]
\includegraphics[width=0.95\linewidth]{"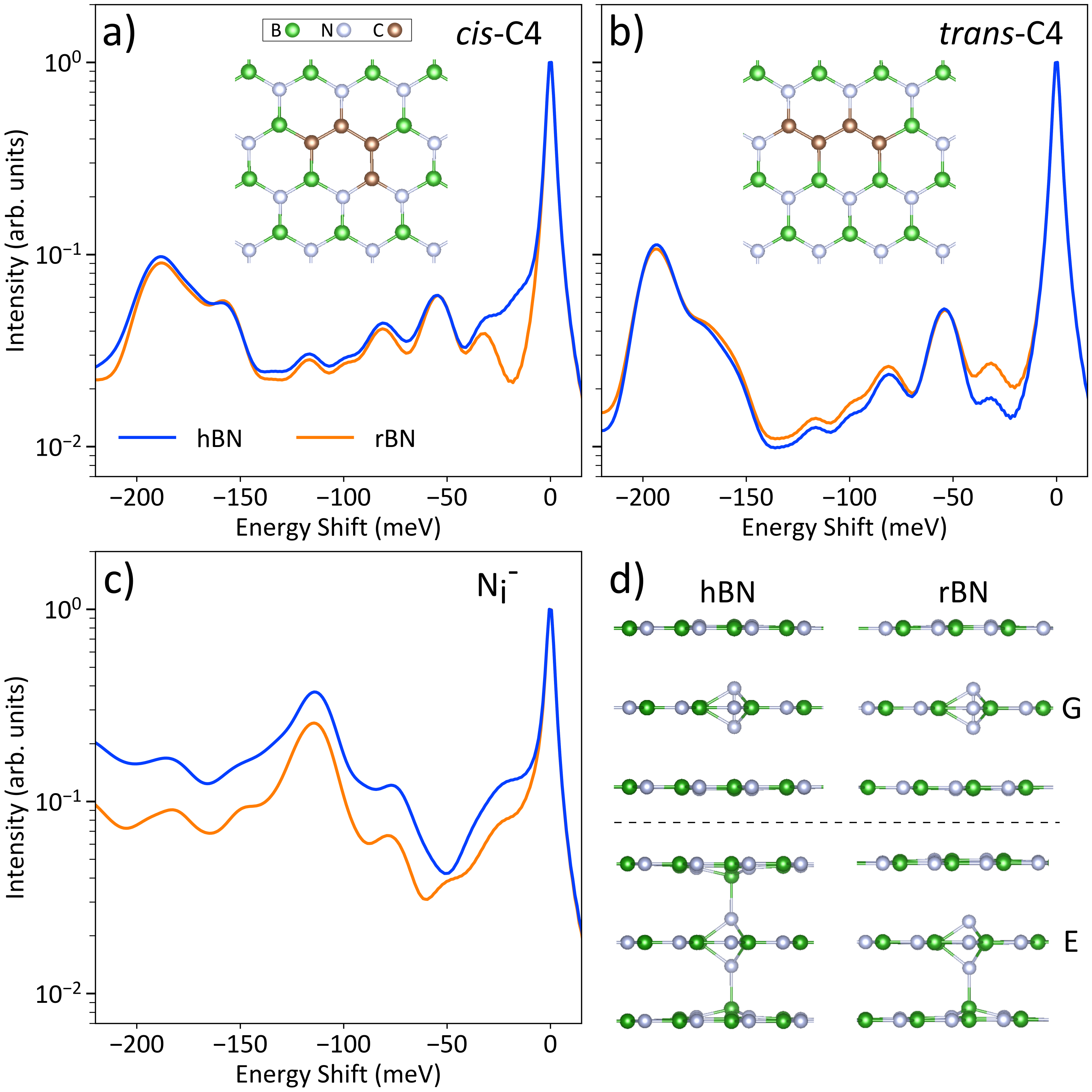"}
\caption{Overview of proposed B-center defect structures and calculated emission spectra in hBN and rBN polytypes. a)-b) Calculated spectra for the in-plane $cis$-C4 (a) and $trans$-C4 (b) defects in each polytype, with the corresponding defect structures shown as insets. c) Calculated spectra for the negatively charged nitrogen split interstitial defect \ce{N_i-}. All spectra are normalized and the ZPL is shifted to 0~meV. d) Defect structures of the \ce{N_i-} defect in each polytype, showing both the ground (G) and excited (E) states.}
\label{Fig_Simulations}
\end{figure}

Next, we consider the in-plane carbon chain tetramer C4, which can exist in the two inequivalent $cis$- and $trans$- configurations, as depicted in Figure~\ref{Fig_Simulations}, and the nitrogen interstitial defect. The latter is expected to reside in the negative charge state, \ce{N_i^-}~\cite{Weston2018,Ganyecz:2024}, with a split-interstitial configuration (Fig.~\ref{Fig_Simulations}d). Upon excitation, the interstitial N-N pair distance increases from 1.60~\AA\ to 1.97~\AA\ in hBN, accompanied by a 0.51~\AA\ out-of-plane displacement of the adjacent boron atoms in the top and bottom hBN layers. In rBN, the defect adopts similar ground- and excited-state structures, with the N-N pair distance increasing from 1.60~\AA\ to 1.94~\AA\ in the excited state, and the boron atom directly below the interstitial nitrogen undergoing an out-of-plane distortion of 0.51~\AA\ (see Figure~\ref{Fig_Simulations}d). 

The calculated emission spectra for the C4 and \ce{N_i-} defects in each BN polytype are shown in Figure~\ref{Fig_Simulations}a-c as a function of the relative energy shift from the ZPL. The calculated PL spectra for the C4 defects exhibit prominent phonon replicas, in agreement with the measured peaks at 185 and 155~meV (Figure~\ref{Fig_Cryo}b), although the calculated intensity of the 155~meV peak is underestimated. Overall, the calculated spectra for both the C4 defects reproduce the main experimental features, and the $cis$ configuration is in better agreement. In contrast, the most intense calculated phonon sideband for the \ce{N_i^-} defect is $\sim 115$~meV below the ZPL, in disagreement with the measured spectra (additional DFT simulation data are presented in Figs. S3-5). This sideband stems from a local vibrational mode stretching the N-N bond in the defect. Furthermore, we find that the calculated ZPL red-shifts for \ce{N_i^-} when going from hBN to rBN, which is inconsistent with experiment (see Figure~\ref{Fig_Cryo}a), whereas it blue-shifts for the C4 defects (see Table~S1).

We therefore conclude that the B-center emission does not originate from \ce{N_i^-}, whilst it is consistent with a C4 defect. The C4 $trans$ and $cis$ configurations in hBN exhibit ZPL energies of 3.198~eV and 3.483~eV, respectively. The former is consistent with the experimental value {2.843~eV} (Figure~\ref{Fig_Cryo}a), noting the well-known tendency of calculations based on the HSE hybrid functional to overestimate ZPL energies. The calculated blue-shift of the ZPL, going from hBN to rBN, is 34~meV for $trans$-C4, in excellent agreement with the experiment, compared to 94~meV for $cis$-C4. We therefore conclude that, the $trans$-C4 defect structure proposed recently by Maciaszek and Razinkovas~\cite{Maciaszek:2024} is the most likely origin of the B-center emission in hBN and rBN.

\section{Conclusion}
We demonstrated that rBN hosts spin defects and quantum emitters whose analogues are well-known in the hBN polytype. Specifically, focused ion and electron beam irradiation were used to engineer \ce{V_B-} and B-center defects in rBN. The emission energy and optically addressable spin sublevels of \ce{V_B-} remain largely unaffected relative to hBN, and the B-center emission is blue-shifted by 42.5~meV. The emission spectra of potential B-center defect structures were simulated and compared against experiment, indicating that the in-plane $trans$-C4 defect is the most likely candidate. Our study provides a clearer understanding of the B-center atomic structure and opens up new opportunities to harness the nonlinear optical properties of rBN with color centers in monolithic integrated photonics.


\section{Methods}
See the Supporting Information for details on experimental methods and DFT calculations.

\section{Acknowledgments}
We acknowledge financial support from the Australian Research Council (CE200100010, FT220100053, DE220101147, DP240103127), the Office of Naval Research Global (N62909-22-1-2028) and the National Natural Science Foundation of China (52025023, 12427806).
Computational resources were provided by the National Computational Infrastructure and the Pawsey Supercomputing Research Centre through the National Computational Merit Allocation Scheme, and by the UQ Research Computing Centre.

\bibliography{references}

\end{document}